# MOBILE ELEMENTS SCHEDULING FOR PERIODIC SENSOR APPLICATIONS

Bassam A.alqaralleh and Khaled Almi'ani

Al-Hussein Bin Talal University

## ABSTRACT

*In this paper, we investigate the problem of designing the mobile elements tours such that the length of each tour is below a per-determined length and the depth of the multi-hop routing trees bounded by k. The path of the mobile element is designed to visit subset of the nodes (cache points). These cache points store other nodes data. To address this problem, we propose two heuristic-based solutions. Our solutions take into consideration the distribution of the nodes during the establishment of the tour. The results of our experiments indicate that our schemes significantly outperforms the best comparable scheme in the literature.*

## I. INTRODUCTION

Many typical applications of wireless sensor networks (WSNs) considers the process of data collection. This data collection is usually accomplished by wireless transmission of the data (possibly) through multiple hops when sensors are deployed in a hostile or hard-to-access environments. In many cases, energy efficiency has been a major concern in wireless communications because increasing energy expenditure limits the operational lifetime of the network. Furthermore, the exhaustion of the energy sources of the sensors in multi-hop scenarios is non-uniform, as nodes that are close to the sink carry heavier data traffic loads and therefore they are likely to be the first to run out of energy. Once these sensors fail, the sink nodes cannot be reached, and as a result, the network stops working even though the nodes that are located far away from the sink may still have sufficient energy. This is a common problem regardless of which communication protocols are used in the network.

In general, in order to significantly increase the lifetime of the network, Mobile Elements (MEs) [1], [2], [3] have been used because it roams in the network and collects data from sensors via short range communications, therefore, the energy consumption is considerably reduced. Thus, the lifetime of the network increases by avoiding multi-hop communication. The main disadvantage of using this approach is the increased latency of the data collection because the speed of mobile element is typically about $0.1 - 2\ m/s$ [4] [5], which results in extensive traveling time for the ME and, also, delay in gathering the data from sensors.

In practice, the ME tour length is often bounded by a predetermined time deadline, either due to timeliness constraints on the sensor data or due to the limited amount of energy available to the ME itself. A possible solution to this problem is to employ more than one ME; however, this solution is often impractical since the cost of MEs is high, and also it might be useless when some sensors are beyond reach of MEs due to its battery limitations.

To solve this problem, many proposals presented a hybrid approach which has been proposed to be a combination of using multi-hop forwarding and the use of mobile element(s). In this approach, a mobile element visits subset of the nodes that is selected to be caching points. These caching points store the data of the nodes that are not included in the tour of the mobile element. Each caching point transmits its data to the mobile element when it becomes within its transmission range. By adopting such an approach, the mobile elements will be able to collect the data of the entire network without the need for visiting each node physically. Figure 1 shows an example for this hybrid approach, where the mobile element visits the caching points to collect the data of the entire network.

In this direction, we investigate the problem of designing the tour of the mobile element and the data forwarding trees, with the objective of minimizing the depth of forwarding trees. We propose two heuristic-based solutions to address this problem. The first heuristic works by recursively partitioning the network; based on the distribution of the nodes. Then in each partition, the process works to determine the caching points that satisfies the constraints. The second heuristic employs similar steps on tree-structured network. The results of our experiments indicate that our schemes significantly outperforms the best comparable scheme in the literature.

The rest of the paper is organized as follows. Section 2 presents the related work in this research area. Section 3 presents the Problem definition. In Section 4, we present an Integer linear program formulation for the presented problem. In Section 5, we present the details of our algorithmic solutions. In section 6, we extended our heuristics to address the situation, where more than one mobile element is available. Section 7 presents the evaluation. Finally, Section 8 concludes the paper.

## II. RELATED WORK

This section reviews the recent literature that studied the use of mobile element(s) to extend the lifetime of sensor networks. We review three major approaches based on the categorization given in [3].

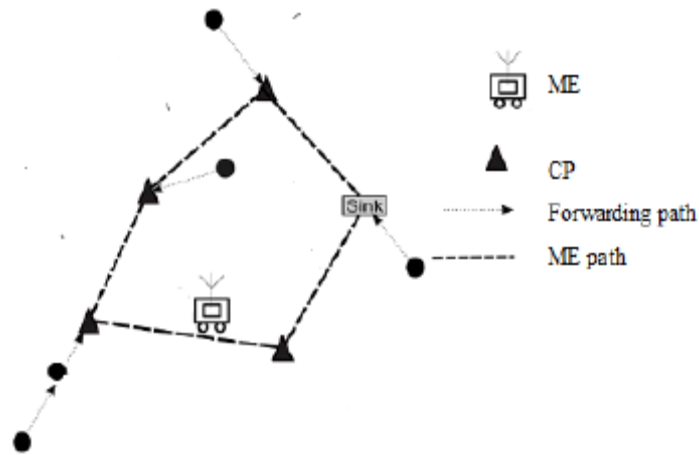

Figure 1: An example showing the mobile element path and the forwarding trees

Firstly, in a typical flat-topology network, the nodes around the sink suffers from heavy load of forwarding the data traffic from all other sensors, consequently, these nodes are likely to be the first to die. Several proposals [6], [7] have investigated the use of mobile sink(s) to reduce the energy consumption in the network. Using this technique, the remaining energy in the nodes becomes more evenly balanced throughout the network by varying the path to the sink(s), leading to a longer lifetime of the network. However, this technique requires the routes to sink node and its location to be changed regularly, which may potentially cause excessive overhead at the nodes due to the frequent re-computation of the routes. Zhao et al [8], [9] presented two distributed algorithms to maximize the overall network utility. These data gathering algorithms are based on placing the mobile sink at each anchor point (gathering point) for a certain period of sojourn time and, on the other hand, mobile sink collects data from nearby sensors via multi-hop communications. They considered both cases of fixed and variable sojourn time.

In the second approach, the paths of the mobile elements are designed to visit each node. In this approach the data of each node is gathered by a mobile element via single-hop communication. In this situation, the path planning problem share fundamental characteristics with the Traveling Salesman Problem (TSP) [10]. It is clear that new constraints must added to capture the characteristics of the sensor environments. In [11], [12], [13], [14], [15], [16], [17], [18], [19] proposed several heuristics to visit each node. This approach provides the capability of significantly reducing the energy consumption by avoiding multi-hop communications, however, it incurs a high delay when the network area is large because the MEs must physically visit all sensor nodes.

Finally, the third approach is a hybrid approach that combines data collection by mobile elements with multi-hop forwarding .Our work is along the line of this approach. Some previous proposals, e.g. [20], [21], [22], assumed that mobile route should be predetermined, and also, were mainly concerned with the timing of transmissions in order to minimize the need for in network caching via timing the transmissions to coincide with the passing of the tour. The problem presented in this share some similarities with minimum-energy Rendezvous Planning Problem (RPP) [23], [24]. In this problem, the objective is to determine the mobile element path such the total Euclidean distance between nodes not included in the tour and the tour is minimized. In [24], the authors presented the Rendezvous Design for Variable Tracks (RD-VT) algorithm. The process of this algorithm starts by construction the Steiner Minimum Tree (SMT) that connects the source nodes. Then, the obtained tree will be traversed in pre-order until no more nodes can be visited without violating the deadline constraint. In this algorithm the visited nodes is identified as the caching points. Xing et al. [23] provided a utility-based algorithm and address the optimal case for restricted version of the problem. Many proposals [25], [26], [27] have also investigated this problem. The problem presented in this work can be categorized as a restricted version of the RPP problem, where the main difference is bounding the depth of the routing trees.

The problem presented in this work also share some similarities with the problem presented by Almi'ani et al. [28]. Their problem deal with designing multiple connected tours such that each node is most k-hope away from one of the tours. The main difference between this problem and our problem is that we focus on the single mobile element case. Also, in the multiple mobile situation, we do not restrict the tours to be connected.

The problem presented in this paper shares some similarities with the Vehicle Routing Problem (VRP) [29]. Given a fleet of vehicles assigned to a depot, VRP deals with the determination of the fleet routes to deliver goods from a depot to customers while minimizing the vehicles' total travel cost.

## III. PROBLEM DEFINITION

We are given an undirected graph $G = (V, E)$, where $V$ is the set of vertices representing the locations of the sensors in the network, and $E$ is the set of edges that represents the communication network topology, i.e. $(v_i, v_j) \in E$ if and only if $v_i$ and $v_j$ are within each others communication range. The complete graph $G' = (V, E')$, where $E' = V \times V$, represents the possible movements of the mobile elements. Each edge $(v_i, v_j) \in E'$ has a length $r_{i,j}$, which represents the time needed by a mobile element to travel between sensor $v_i$ and $v_j$ . The data of all sensors must be uploaded to a mobile element periodically at least once in $L$ time units, where $L$ is determined from the application requirements and the sensors buffer size. In other words, we assume that each mobile element conducts its tour periodically, with $L$ being a constraint on the maximum tour length. In this paper, for simplicity, we assume that the mobile element travels at constant speed, and that, therefore, the travelling times between sensors ($r_{i,j}$) correspond directly to their respective Euclidean distances; however, this assumption is not essential to our algorithms and can be easily dropped if necessary. Also, we are given $k$ that represent the maximum number of hops allowed between any node and its caching point.

In our problem, we seek to find the mobile element tour, where the length of the tour is bounded by $L$, such that the depth of the multi-hop routing trees is minimized and bounded by $k$.

## IV. INTEGER LINEAR PROGRAM FORMULATION

In this section, we present an Integer Linear Program (ILP) for the investigated problem. This ILP is based on the formulation proposed by Almi'ani et al. [26]. We modify their formulation by incorporating the constraint that restrict the number of hops between any node and its caching point to $k$, as an upper bound.

**Variables**

$y_{i,j}$ , $y_{i,j} = 1$ if the edge $(v_i, v_j)$ is included in the ME tour, and 0 otherwise
$x_{i,j}$ , $x_{i,j} = 1$ if the node $v_i$ is included in the tour and is responsible for storing the data of node $v_j$ , which is not included in the tour. $x_{i,j} = 0$ otherwise

**Parameters**

$d(v_i, v_j)$ is the number of hops between $v_i$ and $v_j$ .
$r_{i,j}$ is the travelling time between $v_i$ and $v_j$ .

**Objective**

$$\min \sum_{(i,j)} y_{i,j} \cdot r_{i,j} + M_l \cdot \sum_{(i,j)} x_{i,j} \cdot d(v_i, v_j) \qquad (1)$$

The first term of the objective function is the travelling time of the ME (travel cost), and the second term is the number of hops between the nodes not included in the tour and the tour (assignment cost). The coefficient $M_l$ must be set to any value greater than $max_{i,j} r_{i,j}$. Multiplying the second term by $M_l$ ensures that the assignment cost is strictly prioritized over the travel cost.

**Constraints**

- The load at each node is balanced

$$\sum_{i \in V} y_{i,j} - \sum_{i \in V} y_{j,i} = 0 \qquad \forall j \in V, (2)$$

- the tour starts and ends at vs.

$$\sum_{i \in V} y_{v_s,i} = 1 \quad (3)$$

$$\sum_{i \in V} y_{i,v_s} = 1 \quad (4)$$

- bounds the travelling time of the ME tour.

$$\sum_{i,j \in V} y_{i,j} \cdot r_{i,j} \leq L \quad (5)$$

- Each node must be either involved in the mobile element tour or connected to a node involved in this tour

$$x_{i,j} + \sum_{l \in V} y_{i,l} \leq 1 \quad \forall i, j \in V \, (6)$$

$$\sum_{j \in V} y_{i,j} + \sum_{j \in V} x_{i,j} > 0 \quad \forall i \in V \, (7)$$

- the subtour elimination constraint.

$$z_i - z_j + n \cdot y_{i,j} \leq n - 1 \quad \forall i \in V, \forall j \in V - v_s \, (8)$$

- the number of hops between any node and its caching point is less than or equal to $k$

$$x_{i,j} \cdot d(v_i, v_j) \leq k \quad \forall i, j \in V \, (9)$$

Typically, it is hard to solve such problem. However, the ILP is given to take close look at the problem.

## V. THE ALGORITHMIC SOLUTION

We propose two complementary approaches to address the presented problem. In the first approach, we begin by identifying the best nodes to be used as caching points. Once these nodes are identified, the TSP tour that consists of these nodes is constructed. The last step of this approach is to build the forwarding trees rooted at the caching nodes.

In the second approach, by transforming the network topology into tree-structure, we start by identifying the forwarding trees, and then we determine the caching points. However the first approach leaves more room for optimization for the cache point identification and tour finding steps.

As we will present in this section, both heuristics combine approximation algorithms for two fundamental NP-complete problems: The Dominating Set Problem and The Traveling Salesman Tour Problem. Each of these algorithms is used in different phases of the two heuristics, but in

19

different stages of the algorithm. The design of the heuristics presented in the work is inspired by the P-Based algorithm proposed by Almi'ani et al. [28]. However, as we mentioned before, Almi'ani et al. [28] investigated different problem. Firstly, we describe how these phases work.

A. Selecting the Caching points

Selecting the caching points is the first step in the process of constructing the mobile element tour. The problem we are considering requires that all nodes not included in the tour to be within $k$ hops from their caching points. Therefore, the nodes on the tour will be an Extended Dominating Set of the graph. The nodes outside the tour will need to forward their traffic using multi-hop routing trees that as we know will be bounded in size.

B. Constructing the tour

The tour construction step build the mobile element tour to pass through the caching points identified in the previous step. The objective of the tour building step is to minimize the traveling cost, and therefore it is exactly the TSP problem. Any TSP algorithm or heuristic can be used to obtain the tour of the mobile element. Here, we use the Christofides approximation algorithm, as it is known to behave well in practice.

C. Building the routing trees

Nodes outside the tour will need to forward their data using a multi-hop routing tree. As we will present next, both of the presented heuristics need to solve the k-hop dominating set problem that can be defined as follows:

Given a graph $G = (V, E)$, find a subset of the nodes, such that every node in a graph $G$ is at most $k$ hops away from a node in that subset. The subset is called a k-hop dominating set, and we would like to minimize the size of it. To solve the k-hop dominating set problem, we use the algorithm proposed by Almi'ani et al. [28].

To solve the k-hop dominating set, we first need to construct the graph $G_k$, that will has an edge between any two node, if there is a path between these two nodes in $G$ with at most k-hops. The graph $G_k$ has the same set of vertices as the graph $G$. Now the process iterate until all nodes removed from $G_k$. In each iteration, the node with highest degree (the nodes with highest number of edges) and all of its neighbors will be removed. These removed nodes will be considered as one set. Then, the graph $G_k$ will be altered by removing the edges between the removed node and the nodes that is still in $G_k$. Algorithm 1 shows the steps of this algorithm.

Next we present the Graph Partitioning (GP) and the Tree Partitioning (TP) heuristics.

```
Algorithm 1 DOMINATING SET APPROXIMATION
 1: procedure KDOMAPPROX(G,k)                    ▷ G = (V, E)
 2:     Initialize empty sets D, C
 3:     Initialize G_k
 4:     while G_k has nodes do
 5:         Find v∈G_k that has the highest degree
 6:         Remove v and all of its neighbors from G_k
 7:         Add to the removed nodes to C
 8:         Add C to D
 9:         Empty C
10:     end while
11:     return D
12: end procedure
```

```
Algorithm 2 THE GRAPH PARTITIONING HEURISTIC
 1: procedure GB(G,k, L)
 2:     Initialize empty sets T, R and CPs
 3:     k = 1
 4:     T = L + 1
 5:     while T > L do
 6:         Cps = FindCPS(G,k)
 7:         T = BuildTour(CPs)
 8:         R = BuildRouting(CPs, G)
 9:         k = k+1
10:     end while
11:     return CPs, T and R
12: end procedure
```

## VI. THE GRAPH PARTITIONING HEURISTIC

The goal of this heuristic is to determine the caching points, such that the length of the path between any node and its caching point is within k-hops. In addition, the caching points are selected with the objective of maximizing the lifetime of the network. Once these caching points are obtained, the process proceeds to build the multi-hop routing trees and the mobile element tour.

As we discussed in the previous section, this heuristic consists of three steps, (1) the caching point identification step, (2) the routing trees construction step and (3) the tour building step. At the beginning the $k$ value will be set to one. Once the mobile element tour is obtained, if the obtained tour violates the transit constraint, the process will be repeated and the value of $k$ will be increased by one. Otherwise, the obtained solution will be confirmed as a valid solution. Algorithm 2 shows the steps of this heuristic.

A. Selecting the Caching points

This step starts by solving the k-hop dominating set problem, using the algorithm proposed in the previous section. Solving the k-hop dominating set problem results in group of sets, where in each set there is a node such that the path between this node and any other nodes in the set is below k-hops. Once these sets are identified, the process proceeds to select the candidate caching points from each set. In each set, a node is selected as a candidate caching point, if the path between this node and any other node in the set is at most k-hops.

Then, the process iterates to confirm the final caching points list. In each iteration, the nearest caching point candidate to the confirmed caching points will be selected as a caching point. At the beginning the confirmed caching point list will contain only the sink node. This selected caching point will be added to the confirmed list. All other candidates belong to the same set as the confirmed caching point will be removed from the consideration. The process stops, when each candidate caching point is either confirmed as a caching point or removed from consideration. Algorithm 3 shows the process of this step.

B. The routing trees construction and the tour building steps

Once the caching points set are identified, each node not included in this set will be assigned to its nearest caching point. Then, for each caching point and the nodes assigned to this caching point, a Minimum Spanning Tree (MST) is created to establish the multi-hop forwarding trees.

**Algorithm 3** THE CACHING POINTS IDENTIFICATION STEP

```
1: procedure FINDCPS(G,k)
2:     Initialize empty sets C, D and CPs
3:     D=kDomApprox(G,k)
4:     for each node v in D do
5:         if path between v and all nodes in its sets with k-hops then Add v to C
6:         end if
7:         add sink to CPs
8:     end for
9:     while C has nodes do
10:        Find the closest node in C to any node in CPs
11:        Move this nodes to CPs
12:        Remove any node that belong to same set as this node from C
13:    end while
14:    return CPs,
15: end procedure
```

As we mentioned earlier, the last step is to create the tour of the mobile elements to involve the caching points and the sink. This is established using Christofides algorithm.

## VII. THE TREE PARTITIONING HEURISTIC

The main steps of the tree partitioning heuristic are similar to the steps employed by the graph partitioning heuristic. However, the order in which these steps are used is different in both algorithms.

The tree partitioning heuristic aims to construct its solution by recursively partitioning the routing tree, to identify the caching points and construct the mobile element tour. This heuristic works by first constructing the MST (rooted at the sink) that connect all nodes. By constructing such a tree, the process aims to eliminate costly edges from consideration during the process of obtaining the solution. Then the heuristic proceeds to the caching points identification step. This step is exactly the same as the one in the graph partitioning heuristic. However, the only difference is the type of input, in the TP heuristic, this step uses the MST as an input, where in the GB heuristic, the topology graph is the input. Once the caching points are identified, the routing trees construction and the tour building steps works in the same way as in the GP heuristic. Similar to the GP heuristic, in each iteration, the value of $k$ is incremented by one until a solution that satisfy the transit constraint is obtained.

International Journal of Wireless & Mobile Networks (IJWMN) Vol. 6, No. 1, February 2014

```
Algorithm 4 THE CLUSTERING STEP
1: procedure FINDCLUSTERS(G (topology graph), c (number of clusters to be established))
2:     Randomly choose c nodes as initial cluster centers
3:     while c is not the same in two consecutive iterations do
4:         for each node v in G do
5:             assign each node to the nearest cluster center
6:             assign recalculate center nodes
7:         end for
8:     end while
9:     return Clusters,
10: end procedure
```

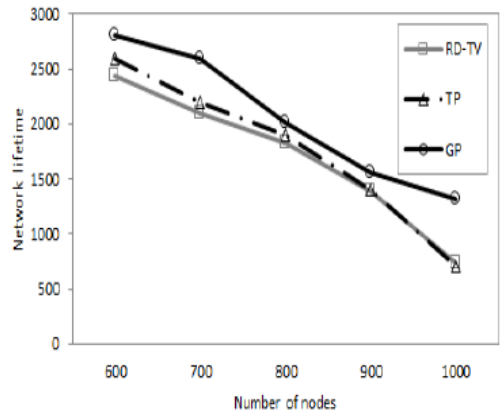

Figure 2: network lifetime against the number of nodes, for the uniform density deployment scenario

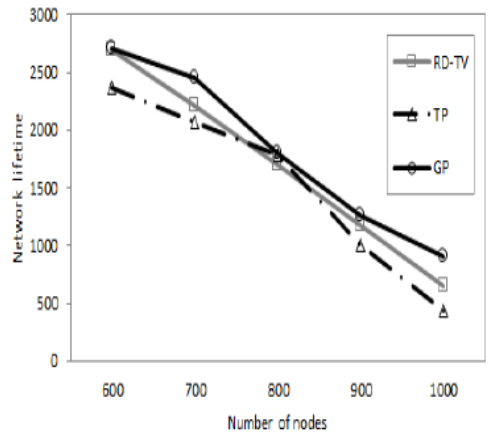

Figure 3: network lifetime against the number of nodes, for the variable density deployment scenario

## VIII. MULTIPLE MOBILE ELEMENTS

In this section, we extended our model to address the situation where more than one mobile element is available (M mobile elements), and therefore we need to design the tours for these mobile element. We assume that the number of available mobile elements is not enough to visit each node.

When multiple mobile elements is available, the main design criteria is how to partition the network based on the available number of mobile elements. The nodes distribution must be considered during the mobile element assignment to ensure that the resultant partitioning will obtain an efficient solution. To this end, we propose the Multi- partitioning (MP) heuristic.

The MP heuristic works by partitioning the network into M partitions, where each partition will be assigned to a mobile element. Then in each partition, the GP heuristic is used to obtain the tour mobile element assigned to this partition.

To obtain the mobile elements partitions, the clustering step is used to obtain M number of clusters. In this step, we use the clustering algorithm proposed by Almi'ani et al. [26]. In this algorithm, the objective is to determine the clusters such that the hop-distances between nodes belong to the same cluster is minimized. This aim to balance the number of nodes belong to the same cluster. Toward this end, the process works by bounding the distance between any node and the center node of this cluster. The center node is defined as the node that is the closest to all other nodes inside the cluster.

Algorithm 4 shows the process of this step. The process starts by selecting c number of nodes randomly as the initial center nodes. Then, each node is assigned to its nearest cluster center node. Once all nodes are assigned, the center node for each cluster is recalculated. The nodes then will be assigned to the new center nodes. The process terminates when the identity of the center nodes does not change between two consecutive iterations.

## IX. SIMULATION METHODOLOGY AND RESULTS

To validate the performance of the presented algorithms, we have conducted an extensive set of experiments using the J-sim simulator for wireless sensor networks [30]. Unless mentioned otherwise, the network area is $250{,}000 m^2$. The value of the tour length constraint $L$ is set to $0.15 \cdot s \cdot T_L$, where $s = 1\ m/s$ is the speed of the mobile element, and $T_L$ is the length of the minimum spanning tree that connects all nodes. The radio parameters are set according to the MICAz data sheet [31], namely: the radio bandwidth is 250 Kbps, the transmission power is $21\ mW$, the receiving power is $15\ mW$, and the initial battery power is 10 Joules. For simplicity, we only account for the radio receiving and transmitting energy. Each node generates one packet in an interval of time equal to $L$. The packet has a fixed size of 100 bytes. Each experiment is an average of 10 different random topologies. We are particularly interested in investigating the following metrics; (1)The lifetime of the network, and (2)Number of caching points. The parameters we consider in our experiments look at varying the number of nodes. We consider the following deployment scenarios:

**Uniform density deployment**: in this scenario, we assume that the nodes are uniformly deployed in a square area of $500 \cdot 500\ m^2$.

**Variable density deployment:** in this scenario, we divide the network into a $10 \cdot 10$ grid of squares, where each square is $500 \cdot 500\ m^2$. We randomly choose 30 of the squares, and in each one of those we fix the node density to be $x$ times the density in the remaining squares. $x$ is a density parameter, which in most experiments (unless mentioned otherwise) is set to $x = 5$.

As a benchmark to the presented algorithms, we compare their performance against the Rendezvous Design for Variable Tracks (RD-VT) algorithm [24]. The RD-VT algorithm works by constructing the Steiner Minimum Tree (SMT) of the source nodes and build a tour based on this tree. To ensure the fairness of the comparisons, we use the Christofides algorithm to find the TSP tour for a given set of nodes in every iteration of the RD-VT algorithm as well. Eventually, each sensor is connected to the nearest point of the tour via the shortest path. With regard to the comparison with the MP heuristic, we adopt the clustering step and run the RD-VT algorithm in each partition.

For simplicity, we only account for the radio receiving and transmitting energy. Figures 2 and 3 show the results for both deployment scenarios as a function of the number of nodes (equivalently, network density). From the figures we can see the GP heuristic outperforms the TP and the RD-VT heuristics in both deployment scenarios. Also, in the uniform deployment scenario, it is clearly shown that the TP heuristic slightly achieve better performance compared to the RD-VT heuristic. In contrast, in the variable deployment scenario, the RD-VT heuristic outperforms the TP heuristic. To understand the factors behind the shown performances, we need to take a close look on the mechanism of each heuristic. In the RD-VT heuristic, the main idea is to traverse the SMT in pre-order, until no more nodes can be visited without violating the transit constraint. The use of such mechanism results in selecting relatively large number of caching points. However, the selected caching points are expected to be very close to each other and the sink. This expected to results in relatively very deep routing trees, since the caching points are not distributed to cover the entire network. The depth of such routing trees must degrade the performance of the RD-VT heuristic, since it is expected to generate a large amount of forwarding traffic. In the TP heuristic, the use of the k-hop dominating algorithm to partition the MST results in introducing dependence between the TP performance and the structure of the MST. The first partition is expected to have the sink node, since it is the node that normally has the highest number of neighbors. Such mechanism is expected to partition the MST into many branches. This is obvious because after the selection of each dominating set the number of tree-branches is expected to increase. The impact of such mechanism is expected to degrade the TP heuristic, especially in the variable deployment scenario, since in this case the MST is expected to have many branches to begin with. In the GP heuristic, the partitioning occurs based on the distribution of the nodes, and therefore this heuristic is expected to have the same performance ratio, regardless of distribution pattern. These factors clarify the shown performances behavior.

Now, we investigate the impact of the number of nodes on the number of caching points that each heuristic obtains. Figures 4 and 5 show the results for both deployment scenarios. From the figures we can see that in both deployment scenarios, the RD-VT heuristic obtains the highest number of caching points. Also, we can see that the GP heuristic consistently obtains the lowest number of caching points. As we mentioned, the RD-VT heuristic results on selecting caching points very close to each other in term of distance. This is the main key behind the RD-VT heuristic capability of obtaining a high number of caching points. In the GP heuristic, the partitioning step has more control over the distribution of the caching points compared to the TP heuristic. This is clearly due to the number of cuts to the original MST tree, after each partitioning step, in the TP heuristic.

Now, we move to compare the performance of the MP heuristic against the modified version of the RD-VT heuristic. Figures 6 and 7 show the impact of the number of nodes on the lifetime of the networks; for both deployment scenarios. From the figures we can see that in both deployment scenarios, the MP heuristic consistently outperforms the RD-VT heuristic. This is expected, since both heuristics use to same strategy to assign the mobile elements to networks partitions, and therefore this comparison must behave similar to one between the GP and the RD-VT heuristics. Figures 8 and 9 show the impact of number of nodes on the number of caching points each algorithm obtains; for both deployment scenarios. Also, this experiment behavior is similar to the one mobile element experiment because of the same mentioned reasons.

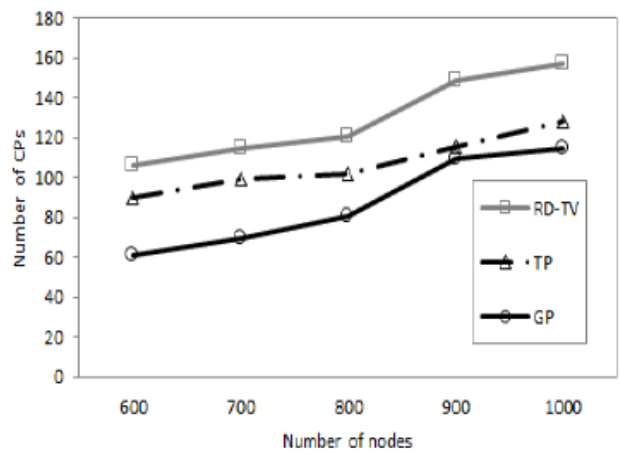

Figure 4: Number of caching points against the number of nodes, for the uniform density deployment scenario

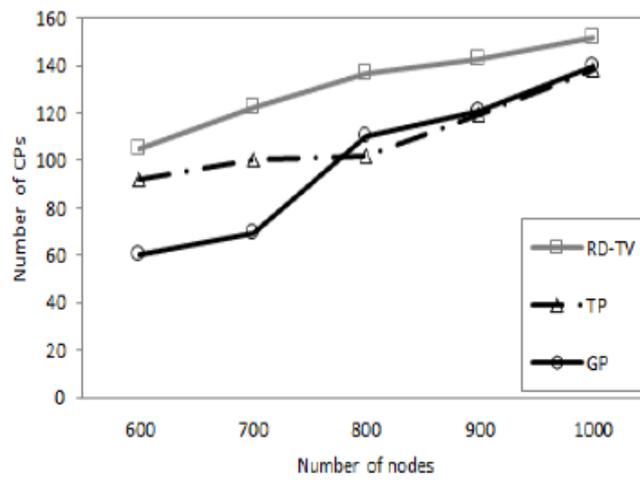

Figure 5: Number of caching points against the number of nodes, for the variable density deployment scenario

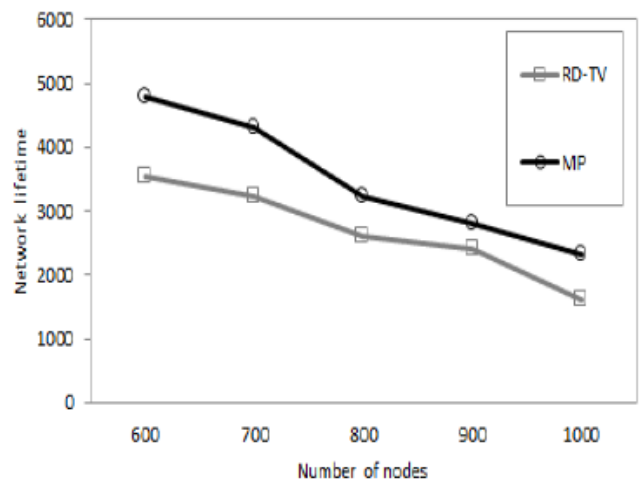

Figure 6: network lifetime against the number of nodes, for the uniform density deployment scenario

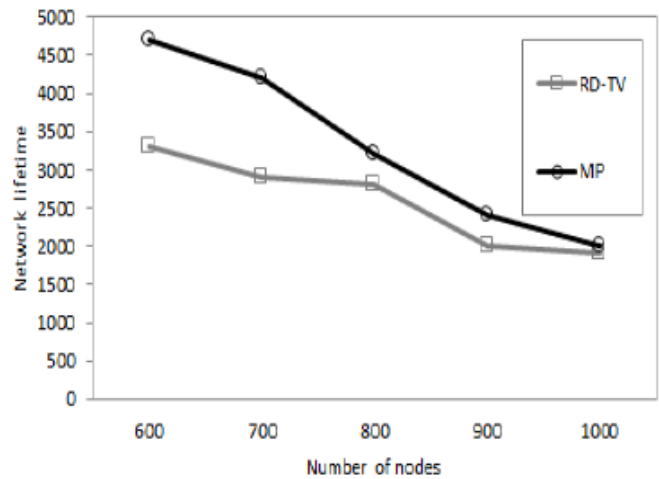

Figure 7: network lifetime against the number of nodes, for the variable density deployment scenario

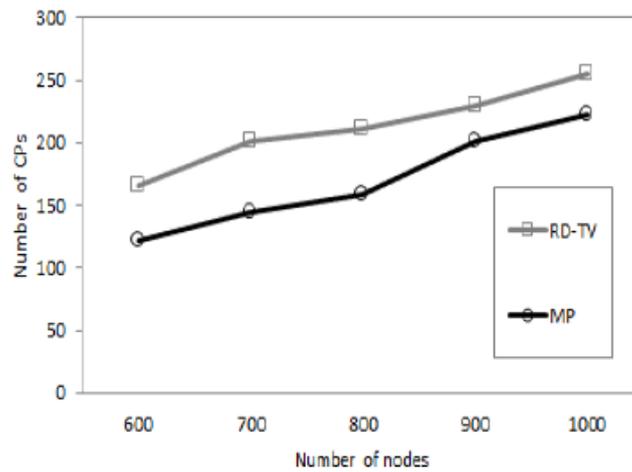

Figure 8: Number of caching points against the number of nodes, for the uniform density deployment scenario

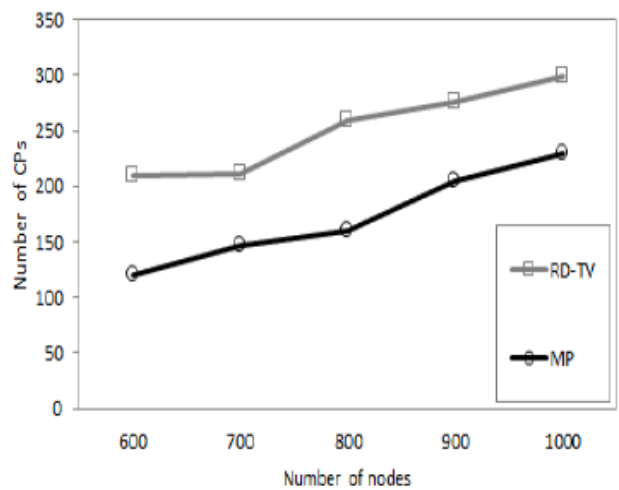

Figure 9: Number of caching points against the number of nodes, for the variable density deployment scenario

## X. CONCLUSIONS

In this paper, we consider the problem of designing the mobile elements tours such that total size of the routing trees is minimized. In this work, we present an algorithmic solutions that creates its solution by partitioning the network, then in each partition; a caching node is selected based on the distribution of the nodes. An interesting open problem would be to consider application scenarios where the data gathering latency requirements vary in the network. For example, some areas in the network need to send data more frequently than others. In this case the tour length constraints would be different for different areas.

**AUTHORS**

Bassam A. Y. Alqaralleh received his BSc degree in Computer Science in 1992, graduate Diploma in Computer Information Systems in 2002 and Masters Degree in Computer Science in 2004. After that, he received Ph. D. in Computer Science from University of Sydney in 2010. His research interests are in the areas of Distributed Systems, Load-Balancing, Networking and Security Systems. He has published a number of conference and journal papers. Currently, he is an assistant professor at the Computer Science Department / Faculty of Information Technology - Al-Hussein Bin Talal University. Since 2013, he is the dean for the Faculty of Information Technology.
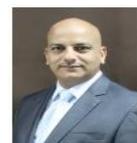

Khaled Almiani is currently an assistant professor at the school of computer Science, Al-Hussein Bin Talal University, Jordan. His main research interests include designing efficient algorithms to improve the performance of WSNs and game-theoretical modeling for WSNs. Khaled received his PhD degree in Information technology from the University of Sydney in 2010.

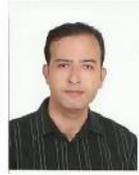